\documentclass[aps,prl,showpacs,twocolumn]{revtex4}
\usepackage{amsmath,amssymb}
\usepackage{graphicx}
\usepackage{color}

\newcommand\xx[1]{\bgroup #1\egroup}

\begin{document}

\title{Melting a granular glass by cooling}

\author{Jan Plagge}\author{Claus Heussinger} \affiliation{Institute for Theoretical Physics,
  Georg-August University of G\"ottingen, Friedrich-Hund Platz 1, 37077
  G\"ottingen}

\begin{abstract}
Driven granular systems readily form glassy phases at high particle volume
fractions and low driving amplitudes. We use computer simulations of a driven
granular glass to evidence a re-entrance melting transition into a fluid state,
which, contrary to intuition, occurs by \emph{reducing} the amplitude of the
driving. This transition is accompanied by anomalous particle dynamics and
super-diffusive behavior on intermediate time-scales. We highlight the special
role played by frictional interactions, which help particles to escape their
glassy cages. Such an effect is in striking contrast to what friction is
expected to do: reduce particle mobility by making them stick.
\end{abstract}

\pacs{} \date{\today}

\maketitle


Friction implies stability.  A solid block only remains static on an inclined
plane because there is friction with the underlying surface.  Similarly, a heap
made of granular particles, in general, will have a higher angle of stability
with increased friction coefficient~\cite{nedderman.book}. Inside the heap, the
particle volume fraction will generally be lower 
and the number of inter-particle contacts smaller, while the heap nevertheless
remains stable~\cite{PhysRevE.65.031304}. A similar stabilization occurs when
the grains are driven into a fluid state.  Under shear, the jamming transition
from a freely flowing state to a yield-stress fluid occurs at lower volume
fractions as compared to the frictionless
case~\cite{bi11:_jamming_by_shear,PhysRevE.83.051301,PhysRevE.84.041308}.

Here, we present simulations of a driven granular system were friction acts
opposite to what is expected from these simple examples.  We show how friction
can lead to anomalous particle dynamics that very efficiently fluidize the
system. As a result the system undergoes a re-entrance melting transition from a
glassy to a fluid state by \emph{lowering} the amplitude of driving.

Experimentally, a variety of driving mechanisms have been proposed to
characterize the dynamical properties of dense granular systems. Among those are
shaking~\cite{Mullin, Yeomans} fluid- or
air-flow~\cite{schroeter2005PRE,keysNatPhys2007,B926754A}, cyclic
shear~\cite{pouliquen2003PRL,dauchot2005PRL} or temperature
oscillations~\cite{chen06:_granul,PhysRevLett.101.148303}. We describe a
two-dimensional system, similar to the setup used in
Refs.~\cite{Mullin,lechenault,B926754A,keysNatPhys2007}. At high densities and
low enough driving amplitude these systems readily form glassy states, where
structural relaxation is completely suppressed~\cite{PhysRevLett.104.225701}.
Interestingly, Ref.~\cite{lechenault} also reports anomalous particle dynamics
deep in the glassy phase, and suspects friction to play a central role in this
process.  Unfortunately, it is rather difficult experimentally to quantitatively
characterize or tune the frictional interactions between the
particles~\cite{PhysRevE.84.031306} or between particles and container walls.
The connection between friction and particle dynamics is therefore unclear.  Our
simulations have the goal to elucidate such a connection.

{\it Model~--~} We consider a monolayer
of $N=2500$ disks.  One half of the particles (``small'') have radius $R_s=0.5d$
and mass $m_s=\rho(4\pi/3)R_s^3$, the other half (``large'') have radius
$R_l=0.7d$ and mass $m_l=\rho(4\pi/3)R_l^3$. The particle area fraction is
defined as $\phi = \sum_{i=1}^N \pi R_i^2/L^2$, where $L$ is the size of the
simulation box. Unless otherwise stated we fix $\phi=0.825$, which is only
slightly below the random close packing value of $\phi_c=0.84$.  Periodic
boundary conditions are used in both directions.

Particles interact via a standard spring-dashpot interaction
(e.g.~\cite{PhysRevE.65.031304,PhysRevE.83.051301,PhysRevE.84.041308}).  In
short, two particles $i,j$ interact when they are in contact, i.e. when their
mutual distance $r$ is smaller than the sum of their radii $R_i+R_j$. The
interaction force has both a normal component $F_n=k_n(r-(R_i+R_j)) $ and a
tangential component $F_t=k_t\delta_t$, where $k_n$ and $k_t$ are the spring
constants and $\delta_t$ is the tangential (shear) displacement since the
formation of the contact. The tangential spring mimics sticking of the two
particles due to dry friction.  The frictional forces are limited by the Coulomb
condition $F_t\leq \mu F_n$, which is implemented by rescaling the tangential
displacement $\delta_t\to \mu F_n/k_t$ whenever necessary. 

To mimic the presence of an external container, we place the particles on a flat
surface ($xy$-plane) \xx{which acts like a frictional particle with infinite
  radius.  The normal component of the interaction is set by gravity, $F^{(s)}_n
  = m_ig$, which pushes the particles into the surface. Therefore, and because
  we do not allow particles to move away from the surface ($z$-direction),
  particles are in permanent contact with the surface.  The tangential
  displacement $\delta_t$ then corresponds simply to the in-plane displacement
  of the particle, properly rescaled when the Coulomb condition is violated,
  $\delta_t \leq\mu_s m_ig/k_t$, where $\mu_s$ is the surface friction
  coefficient.}


Particles are driven with an oscillating force $F(t)=A\sin(\omega t)$ that acts
(in the plane of the surface) along the y-direction. Even though this driving is
uni-directional, we find that due to the dense packing the system remains
roughly isotropic.
With this kind of driving force, an isolated particle on a frictionless surface
oscillates at an amplitude $y_0 = A/m\omega^2$. In a dense assembly this leads
to local frustration as particles with smaller masses tend to move faster.
Rearrangements result which, at the high densities under consideration, may or
may not be able to lead to structural relaxation. It is this glassy dynamics
that we are interested in, with the driving amplitude $A$ playing the role of
thermal temperature. Note, that this driving mechanism injects energy directly
into the bulk.  System-size is therefore not an issue and the system is
spatially homogeneous.

\xx{As units we choose particle mass density $\rho$, particle diameter $d$ and
  the period of the driving, $T=2\pi/\omega$.  With these definitions we perform
  molecular dynamics simulations using LAMMPS~\cite{lammps} with parameters
  $k_n=1000$, $k_t=2k_n/7$ and a time-step of $\Delta t=0.001$.}

The simplest quantity which is measured from particle displacements is the
mean-squared displacement \xx{(MSD),
\begin{eqnarray}\label{eq:}
\Delta^2 (t) = \left\langle \frac{1}{N}
  \sum_{i=1}^N [x_i(t_0+t) - x_i(t_0)]^2 \right\rangle
\end{eqnarray}
where $\Delta x_i(t_0,t) = x_i(t+t_0)-x_i(t_0)$ is the displacement of particle
$i$ in the time interval $[t_0,t_0+t]$} in the direction transverse to the
driving. Snapshots of the system are taken after every full force cycle. Time is
therefore restricted to $t \equiv t_n = nT$ and $n$ integer. With this
definition the MSD is zero when particle motion during cycles is periodic.

\begin{figure}[t]
 \begin{center}
   \includegraphics[width=0.8\columnwidth]{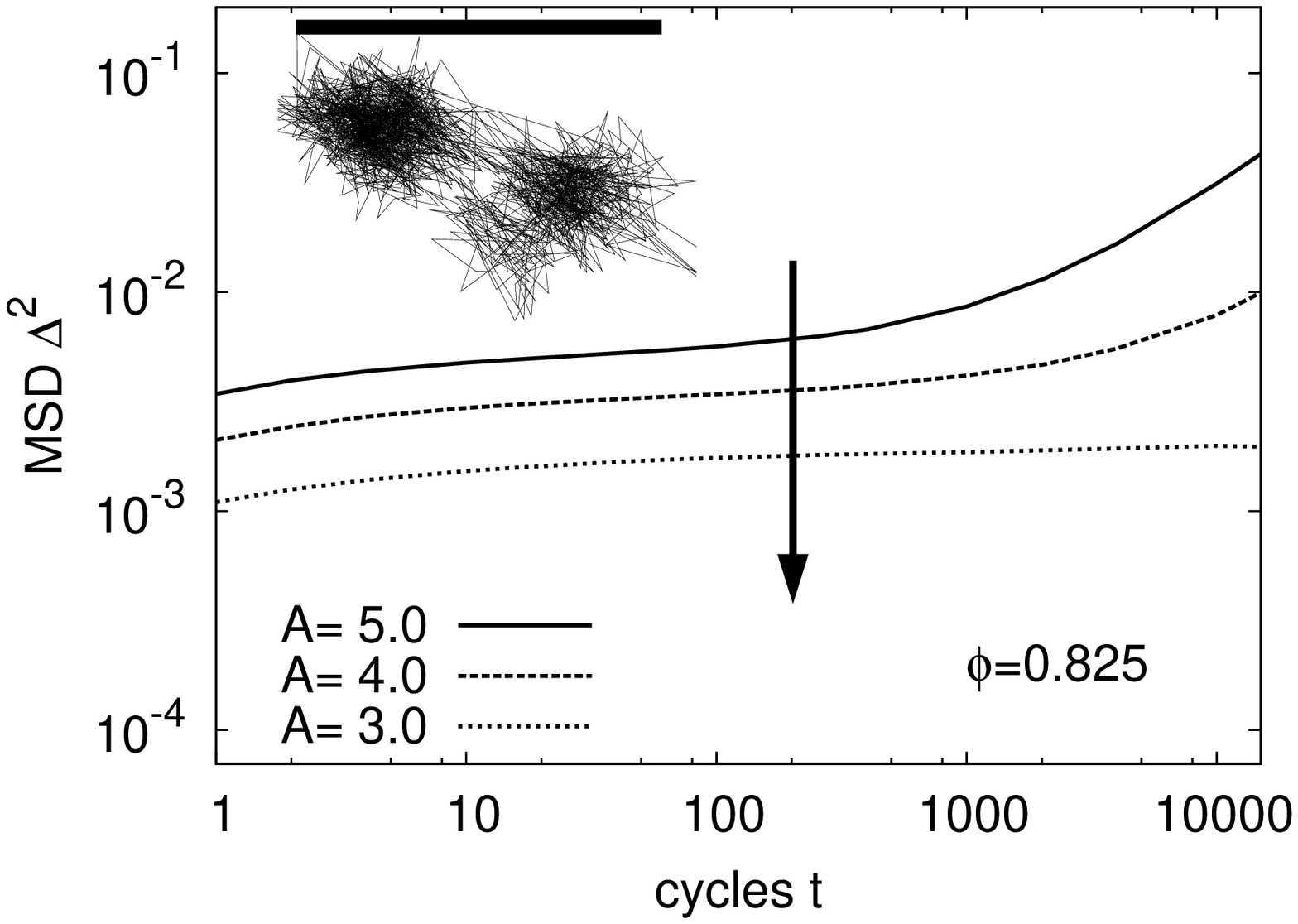}
   \includegraphics[width=0.8\columnwidth]{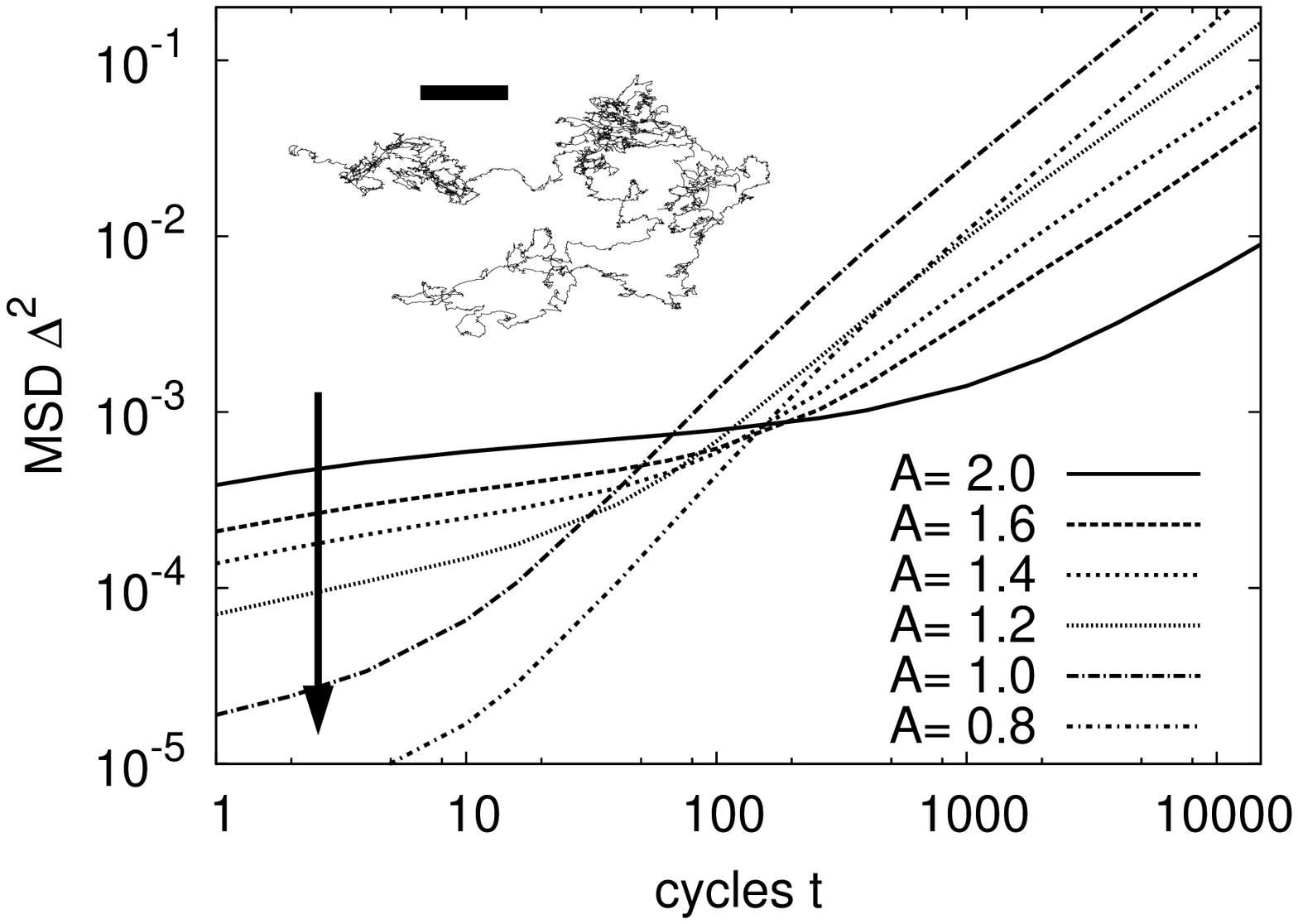}
\end{center}
\caption{Mean-square displacement $\Delta^2(t)$ for different driving amplitudes
  $A$ (arrows indicate decreasing amplitude) .  At short times the plateau
  decreases with decreasing amplitude.  By way of contrast, at long times the
  MSD is nonmonotonic with $A$ and has a maximum at $A^\star\approx 1$. On
  intermediate time-scales superdiffusive particle motion develops when
  $A\approx A^\star$.  \xx{Inset: typical trajectories in glassy regime (top)
    and in the re-entrance fluid (bottom). Scale-bar is of length $R_s$.}  }
  \label{fig:msd}
\end{figure}

{\it Results~--~} Let us first consider the case where there is no
inter-particle friction ($\mu=0$) and only particle-surface friction
($\mu_s=1$). In Fig.~\ref{fig:msd} we display the evolution of the MSD for
various driving amplitudes $A$.

\begin{figure}[t]
 \begin{center}
   \includegraphics[width=0.8\columnwidth]{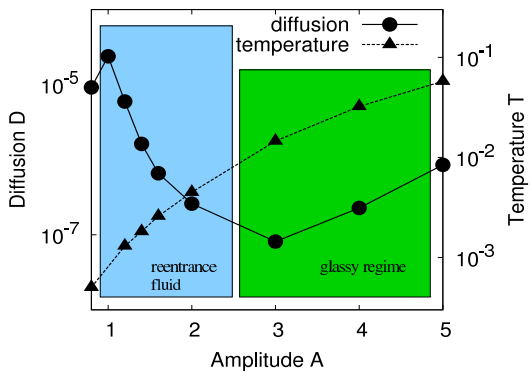}
\end{center}
\caption{ The effective diffusivity $D$ is nonmonotonous with a minimum in the
  glassy regime and a maximum at $A^\star\approx 1$. The temperature $T$ is
  monotonously decreasing with decreasing amplitude.}
 \label{fig:msd.slope}
\end{figure}

At relatively high amplitudes ($A\geq 3$) we observe typical signs of glassy
dynamics; an extended plateau in the MSD, caging of particles and hopping events
\xx{(see sample trajectory in Fig.~\ref{fig:msd}a inset)}. As expected the MSD
decreases with decreasing driving amplitude. Surprisingly, this trend does not
persist when the amplitude is further reduced.  Instead, for lower amplitudes we
observe a greatly enhanced MSD and a total dissapearance of the plateau.  Thus,
the system is fluidized, very unintuitively, by reducing the amplitude of
agitation (also see suppl. movies). This re-entrance melting transition is our
main result.

The fluidization is particularly evident in the long-time diffusivity $D :=
\lim_{t\to\infty}\Delta^2(t)/t$ (see Fig.\ref{fig:msd.slope}), which has a
minimum in the glassy regime~\footnote{In this regime we cannot reach the
  long-time diffusive regime and $D$ is only an upper bound for the real
  diffusivity.} but then strongly increases up to a sharp maximum at
$A^\star\approx 1$. At the same time, the ``granular temperature'' $T$, defined
as the average kinetic energy in the direction transverse to the drive, is
completely normal and monotonously decreases with decreasing $A$.  This
parallels the behavior of the MSD at short times. Anomalous dynamics only
develops at intermediate time-scales. Associated with the maximum of
$D(A^\star)$, we observe an intermediate super-diffusive regime that spans one
to two orders of magnitude in time.





\begin{figure}[t]
 \begin{center}
   \includegraphics[width=0.475\columnwidth]{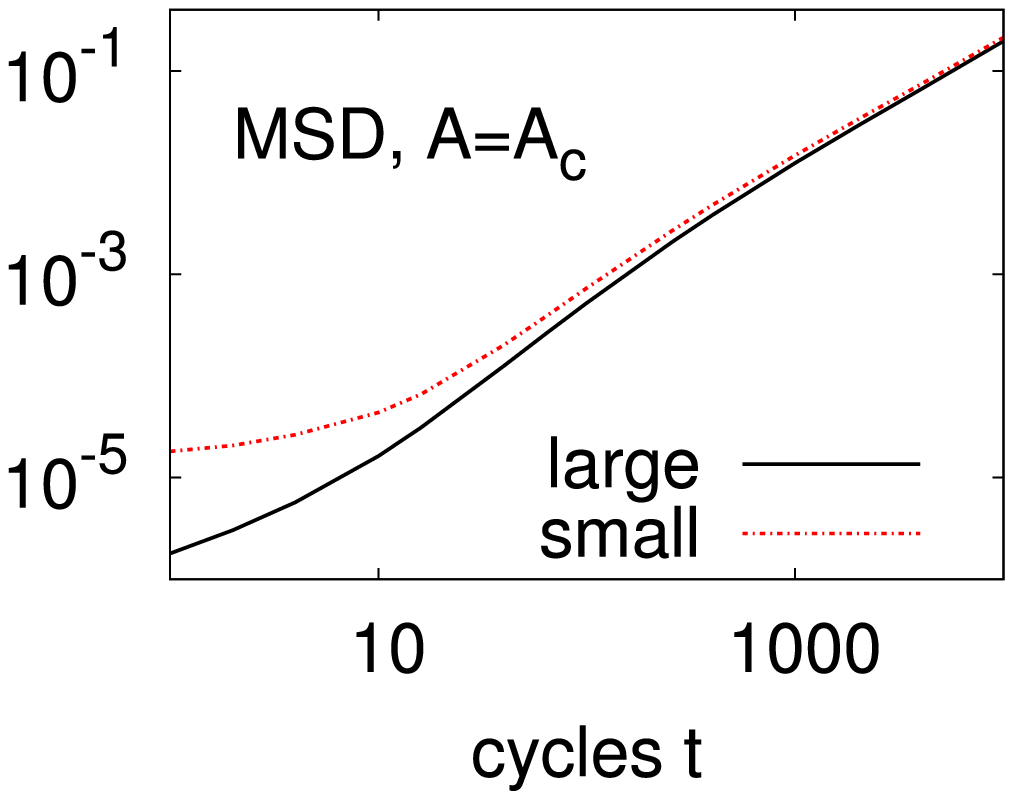}
   \includegraphics[width=0.49\columnwidth]{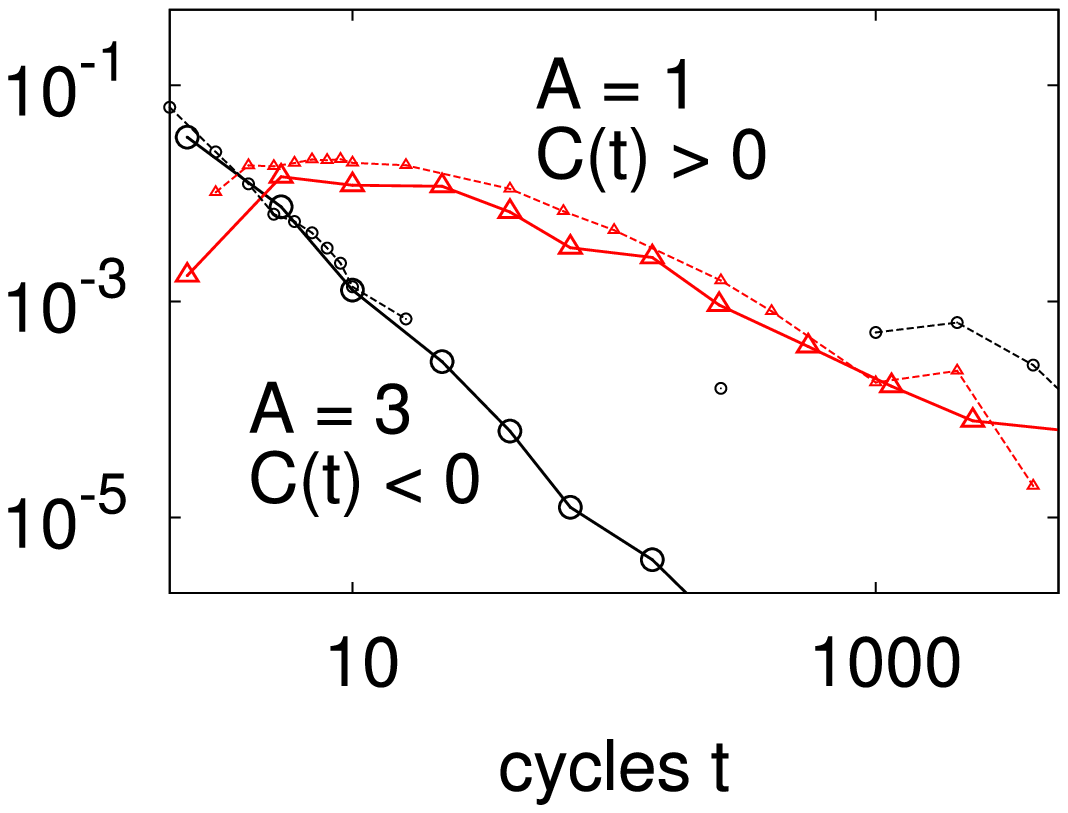}
\end{center}
\caption{ (left) MSD of large particles is similar to that of small particles,
  except at short times, where large particles move much less. (right)
  \xx{Normalized autocorrelation function $C(\tau)/C(0)$ of single-cycle
  displacements.  Negative correlations in glassy regime ($A=3$, plotted is
  $-C(\tau)$) compare with positive correlations in re-entrance fluid ($A=1$).
  Data taken by direct evaluation of the correlations (small symbols) and by
  differentiation of MSD, via $C(\tau)=\partial^2\Delta^2(\tau)/\partial\tau^2$
  (large symbols).}  }
  \label{fig:light.heavy}
\end{figure}

{\it Discussion~--~} \xx{ The external driving injects momentum into the system.
  This momentum is subsequently randomized by inter-particle collisions, and
  dissipated by surface friction. The onset of re-entrance melting corresponds
  to the situation that dissipation of momentum competes with the randomization
  due to collisions. This can most easily be seen in a system, where only the
  small particles are driven. The large particles then only move because they
  are kicked around by the mobilized small particles. We can show (see suppl.
  material) that in the re-entrance fluid phase kicks only temporarily mobilize
  the large particles.  They then undergo some small slip displacement and
  quickly come to rest before the next collision occurs. Thus, all the momentum
  from the collision is immediately lost to the surface.  By way of contrast, in
  the glassy phase this momentum is first redistributed to other particles
  before it is dissipated away.

  Similar effects occur when both particles are driven. The value of $A^\star=1$
  is below the Coulomb threshold for the large particles, which in our units is
  at a force $A_l=1.43$. This means that large particles can only be mobilized
  by additional kicks from mobile small particles, which have a smaller Coulomb
  threshold of $A_s=0.52$.  
  Accordingly, the MSD of large particles is suppressed on short times
  (Fig.\ref{fig:light.heavy}a) but otherwise displays the same superdiffusive
  behavior on intermediate time-scales.  }

\begin{figure}[t]
 \begin{center}
   \includegraphics[width=0.45\columnwidth]{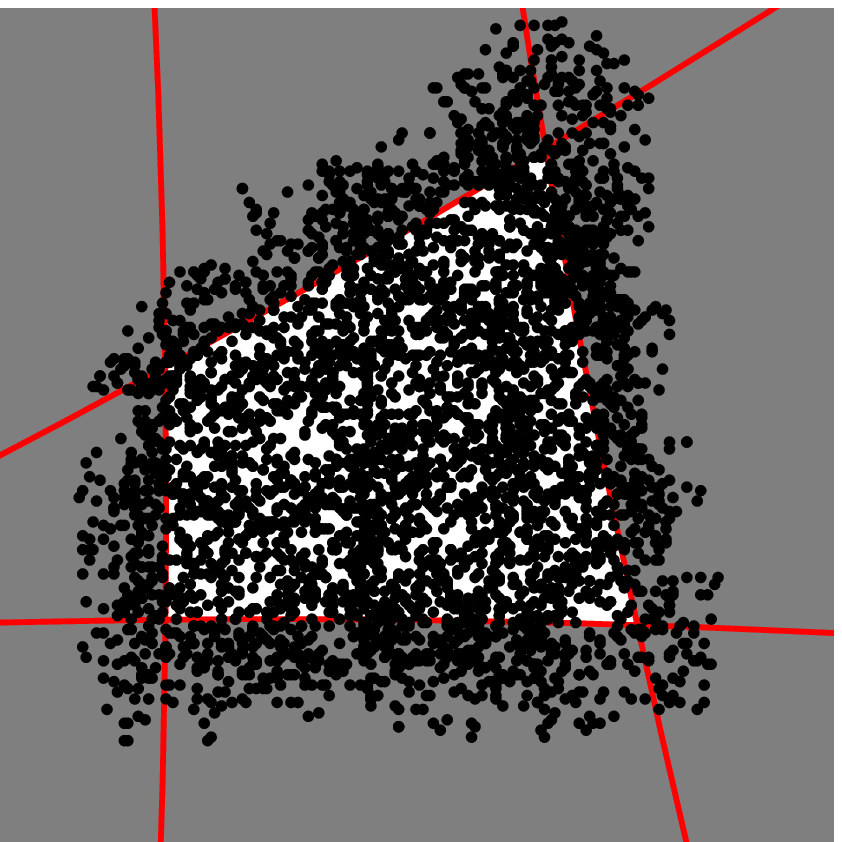}
   \includegraphics[width=0.45\columnwidth]{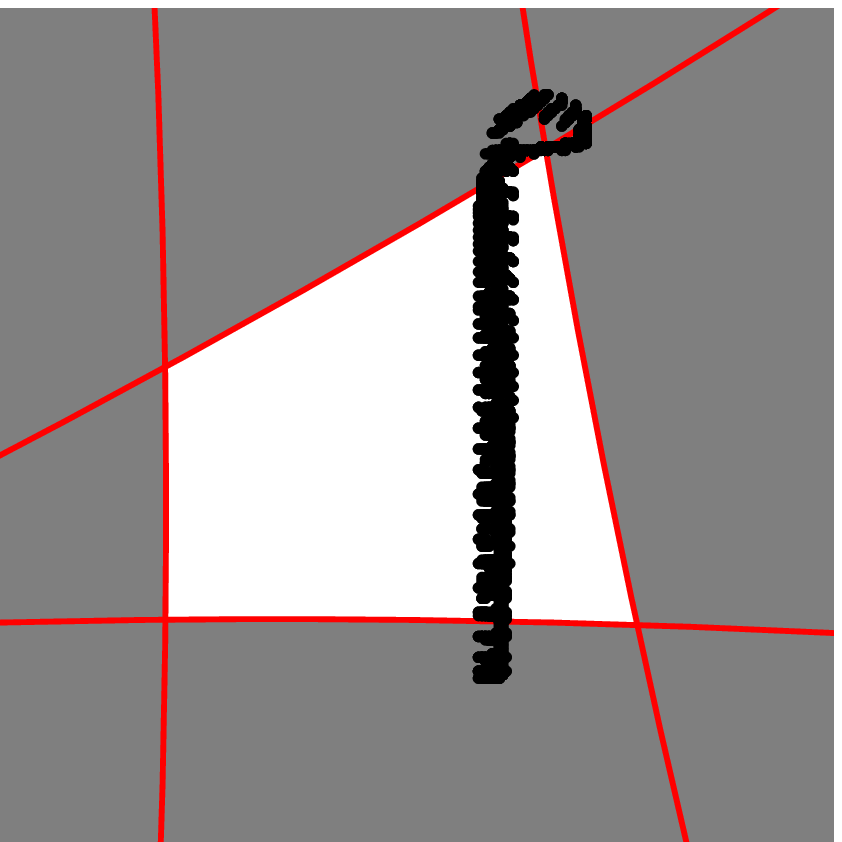}
\end{center}
\caption{ \xx{Short-time ($t=30$) motion of a small test particle (black dots)
    as driven by the periodic force without (left) and with surface friction
    (right).  The test particle is confined by a set of large particles, which
    are frozen in space.  The ``excluded'' space (grey area) is formed from the
    area covered by ``effective'' particles of radius $R_s+R_l$. For the high
    densities under consideration, the available space for particle motion
    (white area) is much smaller than the particle diameter. The boundaries
    therefore appear on this scale only with a small curvature (red lines). As
    the particles are not infinitely stiff, some overlap during collisions
    (penetration into the grey area) is allowed.}}
  \label{fig:traj.cage}
\end{figure}

\xx{ Superdiffusion then naturally emerges, when there is positive temporal
  correlations in the kicks. Fig.~\ref{fig:light.heavy} displays the
  auto-correlation function $C(t)=\langle\Delta x(t_0,1)\Delta
  x(t_0+t,1)\rangle$ of large-particle displacements, $\Delta x(t,1)$, during a
  single cycle.  Clearly, a pronounced positive correlation is visible at
  intermediate times corresponding to the superdiffusive regime. The same
  positive correlations are visible in the trajectory displayed in Fig.1b
  (inset) where they lead to long stretches of quasi-directed motion. By way of
  contrast, the correlation function is negative in the glassy regime,
  indicating anti-correlations in the kicks.

  To understand the origin of these correlations, we have to analyze in more
  detail the dynamics of the small particles. With the large particles sticking
  to the surface, the small particles explore their local free volume on short
  times within an effectively \emph{frozen} environment. This situation is
  depicted in Fig.\ref{fig:traj.cage}, where a small test particle (black dots)
  is confined to a typical cage-like surrounding, which is taken to consist of
  large particles that are frozen in space (grey area and lines).  Driven by the
  external force the test particle will move around and explore the available
  free volume (white area).  }

If we switch off the frictional interactions of the test particle with the
underlying surface (Fig.\ref{fig:traj.cage}a), the motion is quickly randomized
by the collisions with the cage.  \xx{This builds up a pore pressure, which can
  act as a restoring force when, on longer time-scales, the large particles
  (cage wall) start to move. The consequences are anticorrelation (negative
  $C(\tau)$), particle localization and caging.  

}

With friction turned on (Fig.\ref{fig:traj.cage}b), the trajectory is completely
different and the test particle explores only a small part of the available
volume.  \xx{Any additional momentum from an inter-particle collision is quickly
  dissipated and the particles stick to the surface for as long as the force
  needs to switch sign and push it back. This allows to synchronize with the
  force and avoids the chaotic type of motion characteristic of the glassy
  regime.}
The synchronization is apparent in the phase of the oscillations of the
particle. Without friction the particle coordinate is out-of phase with the
forcing, just like a driven and undamped oscillator. With friction the particle
velocity is in phase with the forcing, like an overdamped oscillator (see suppl.
material).

\xx{ On intermediate time-scales the large particles move and the cage evolves.
  If the kicks are sufficiently weak (small $A$) the cage will only change very
  little from one cycle to the next and the periodic trajectories of the small
  particles slowly evolve with the surrounding structure.
  On this intermediate time-scale the kicks are correlated and lead to
  superdiffusive motion of the confining large particles (and as a consequence
  also of the confined small particles). With increasing driving amplitude the
  kicks get stronger and stronger, and the positive correlation is reduced.
  Finally, at high enough amplitudes the large particles are fully mobilized and
  cannot dissipate their momentum before the next kick occurs. This completely
  randomizes particle motion and corresponds to the transition into the glassy
  regime.  }

\begin{figure}[t]
 \begin{center}
   \includegraphics[width=0.8\columnwidth]{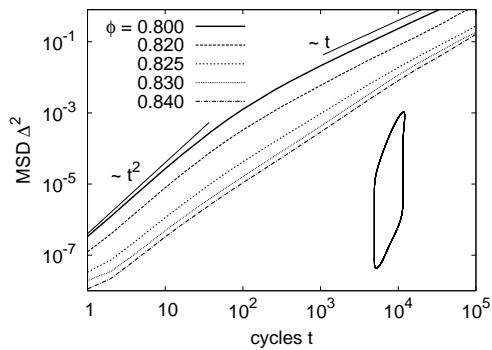}
 \end{center}
 \caption{MSD with interparticle friction and strong damping $\zeta=50$ chosen
   such that velocities are damped on time-scales shorter than the driving
   period.  The driving force ($A=50$) is only applied to small particles. If
   all particles were driven, no rearrangements would occur (overdamped limit).
   \xx{The MSD decreases upon increasing the volume-fraction towards the
     random-close packing limit.}  Inset: example of a periodic particle
   trajectory when friction is switched off completely.  The amplitude of the
   motion corresponds to roughly $5\%$ of the particle diameter.}
  \label{fig:msd.inter}
\end{figure}

{\it Inter-particle friction~--~}
\xx{The above discussion of the origin of anomalous diffusion shows that surface
  friction essentially acts as a strong dissipation mechanism.  However, if we
  substitute surface friction with a strong \emph{linear} damping force, $\vec
  F_{\rm damp}=-\zeta \vec v_i$, and in the absence of inter-particle friction
  ($\mu=0$), no anomalous dynamics occurs.  In fact, the particles move on
  strictly periodic trajectories (Fig.~\ref{fig:msd.inter} inset) and the MSD is
  identical zero. Such a behavior parallels the caging dynamics seen in the
  glassy regime, with the effective cage shrinking to a point.  This suggests
  that the \emph{nonlinear} nature of friction is also essential for the
  anomalous dynamics. Indeed, if we switch on inter-particle friction ($\mu=1$),
  particles readily diffuse around and the MSD displays again a super-diffusive
  (or even ballistic) regime at short and intermediate time-scales
  (Fig.~\ref{fig:msd.inter}). The second role of friction, next to dissipation,
  is thus to induce small perturbations during particle collisions, such that
  the periodic trajectories are slightly, but irreversibly modified. This leads
  to a slow but steady evolution of the local structure, which is visible in the
  MSD as ballistic regime.}

{\it Conclusion~--~}We have shown that, quite unintuitivly, frictional
interactions can speed up particle motion and lead to superdiffusive dynamics.
This contrasts with what friction is expected to do: slow down particle motion
by making them stick. We have rationalized this nontrivial behavior by
considering friction as a weak irreversible perturbation to genuinely periodic
trajectories. This fluidization competes with the localization due to randomized
particle motion.  As a result we encounter a re-entrance melting transition from
a granular glass (localized) to a fluid state upon \emph{lowering} the
temperature.

These results may be important for experiments in several ways. First, note that
the role of frictional interactions with the surface is quite subtle.
Fluidization only happens on relatively long time-scales, while on short times,
particle motion is completely normal.  There, friction only leads to a
suppression of particle motion because of the Coulomb threshold. Such effects
would make it hard to evaluate the relevance of friction in experiments,
whenever interactions with an external container are to be avoided.

In the experiments of Ref.~\cite{lechenault} superdiffusive particle motion was
indeed observed, however only at volume fractions around the critical close
packing limit.  In contrast, we find anomalous dynamics for a range of densities
(Fig.~\ref{fig:msd.inter}). The high stiffness of the brass particles in the
experiment makes a key difference with the simulation. We speculate that only at
$\phi_c$ this high stiffness provides a sufficiently tight packing, such that
frictional effects can compete with the randomization due to the driving. The
role of friction would then be to provide the nonlinear ingredient that can
``rectify'' the motion.  Our analysis suggests that important additional insight
can be obtained by changing the driving amplitude. First
results~\cite{coulais.preprint} indeed suggest that the anomalous dynamics is
enhanced for lower driving amplitudes.

Finally, the transition between reversible and irreversible particle motion is a
nice example of a transition into an absorbing state, as described in the
experiments of Pine et al.~\cite{pine2005Nature}. Interestingly, in our case the
particles continue to interact in the absorbing state. This can, for example, be
seen from the periodic trajectories of individual particles, which are complex
loops and not just straight lines (inset of Fig.\ref{fig:msd.inter}). It remains
to be seen if a volume fraction can be identified, which plays the role of the
critical point of this non-equilibrium transition. We leave this question for
future work.

\acknowledgments

We acknowledge support by the Deutsche Forschungsgemeinschaft, Emmy Noether
program: He 6322/1-1.


\end{document}